\documentclass{nature}

\usepackage{amssymb}
\usepackage{amsmath}
\usepackage{graphicx}

\newcommand{\sun}{\odot}
\newcommand{\etal}{\textit{et al.}}
\newcommand{\biba}[7]{\bibitem{#1}#2 #3. \textit{#4} \textbf{#5,} #6 (#7).}
\newcommand{\mnras}{Mon. Not. R. Astron. Soc.}
\newcommand{\aj}{Astron. J.}
\newcommand{\aap}{Astron. Astrophys.}
\newcommand{\apj}{Astrophys. J.}
\newcommand{\apjl}{Astrophys. J.}
\newcommand{\apjs}{Astrophys. J. Suppl. Ser.}
\newcommand{\actaa}{Acta Astron.}
\newcommand{\memsai}{Mem. Soc. Astron. Ital.}
\newcommand{\tE}{t_{\rm E}}

\title{No large population of unbound or wide-orbit Jupiter-mass planets}

\author{Przemek Mr\'oz$^{1}$\footnote{Corresponding author.}, Andrzej Udalski$^1$, Jan Skowron$^1$, Rados\l{}aw~Poleski$^{2,1}$, Szymon Koz\l{}owski$^1$, Micha\l{}~K. Szyma\'nski$^1$, Igor Soszy\'nski$^1$, \L{}ukasz Wyrzykowski$^1$, Pawe\l{} Pietrukowicz$^1$, Krzysztof Ulaczyk$^{3,1}$, Dorota Skowron$^1$ \& Micha\l{} Pawlak$^1$}

\begin{document}

\maketitle

\begin{affiliations}
 \item Warsaw University Observatory, Aleje Ujazdowskie 4, 00-478 Warsaw, Poland
 \item Department of Astronomy, Ohio State University, 140 W. 18th Ave., Columbus, OH 43210, USA
 \item Department of Physics, University of Warwick, Coventry CV4 7AL, UK
\end{affiliations}

\begin{abstract}
Planet formation theories predict that some planets may be ejected from their parent systems as result of dynamical interactions and other processes\cite{rasio,marzari,veras}. Unbound planets can also be formed through gravitational collapse, in a similar way to that in which stars form\cite{kroupa2003}. A handful of free-floating planetary-mass objects have been discovered by infrared surveys of young stellar clusters and star-forming regions\cite{zapatero,liu} as well as wide-field surveys\cite{dupuy}, but these surveys are incomplete\cite{scholz,pena,muzic} for objects below $5\ M_{\rm Jup}$. Gravitational microlensing is the only method capable of exploring the entire population of free-floating planets down to Mars-mass objects, because the microlensing signal does not depend on the brightness of the lensing object. A characteristic timescale of microlensing events depends on the mass of the lens: the less massive the lens, the shorter the microlensing event. A previous analysis of 474 microlensing events found an excess of very short events\cite{sumi} (1--2 days) -- more than known stellar populations would suggest -- indicating the existence of a large population of unbound or wide-orbit Jupiter-mass planets (reported to be almost twice as common as main-sequence stars). These results, however, do not match predictions of planet formation theories\cite{veras,mao} and are in conflict with surveys of young clusters\cite{scholz,pena,muzic}. Here we report the analysis of a six times larger sample of microlensing events discovered during the years 2010--2015. Although our survey has very high sensitivity (detection efficiency) to short-timescale (1--2 days) microlensing events, we found no excess of events with timescales in this range, with a $95\%$ upper limit on the frequency of Jupiter-mass free-floating or wide-orbit planets of 0.25 planet per main-sequence star. We detected a few possible ultrashort-timescale events (with timescales of less than 0.5 day), which may indicate the existence of Earth- and super-Earth-mass free-floating planets, as predicted by planet-formation theories. 
\end{abstract}

The sample of 2,617 microlensing events we analysed was selected from data collected during the fourth phase of the Optical Gravitational Lensing Experiment\cite{udalski} (OGLE-IV) during the years 2010--2015. The survey is monitoring dense fields towards the Galactic centre, nine of which (about 12.6 square degrees in total) were observed with a cadence of either 20 min or 60 min, allowing the detection of extremely short microlensing events. We analysed the light curves of almost 50 million stars identified on deep stacked images of each field; each light curve consisted of 4,500--12,000 data points.

The selection of events was conducted in three steps, described in detail in the Methods. First, we searched for ``bumps'' in the light curves, which we define as at least three consecutive points $3\sigma_{\rm base}$ above the baseline level ($\sigma_{\rm base}$ is the dispersion of points outside a 360-day window centred on the bump). To minimize contamination from moving objects (like asteroids) and photometry artifacts, we required that the centre of the additional flux coincided with the centre of the star. Events with a very low signal-to-noise ratio and those exhibiting a variability in the baseline were also rejected by these criteria.
Next, we removed any remaining artefacts located mainly near the edges of the CCD camera, low-amplitude pulsating red giants, and other variable stars (dwarf novae, flaring stars) that have multiple bumps in their light curves.

Finally, we fitted the microlensing point-source point-lens model to the data and required that the model describe the data appropriately. The lensing model has three parameters: the time $t_0$ and the projected separation $u_0$ (in Einstein radius units) between the lens and the source during the closest approach, and the Einstein radius crossing time $t_{\rm E}$. (The angular Einstein radius $\theta_{\rm E}$ of a lens depends on its mass $M$ and relative lens-source parallax $\pi_{\rm rel} = 1\ \mathrm{au} (D_{\rm L}^{-1} - D_{\rm S}^{-1})$, where $D_{\rm L}$ and $D_{\rm S}$ are the distances to the lens and the source, respectively, as follows: $\theta_{\rm E} = \sqrt{\kappa M \pi_{\rm rel}}$, where $\kappa=8.14$ mas/$M_{\odot}$). 
Two additional parameters describe the source flux $F_{\rm s}$ and blended unmagnified flux $F_{\rm b}$ from possible unresolved neighbours and/or the lens itself. To ensure that the shortest events were not mistaken for stellar flares, we required at least four data points on the rising branch of the light curve (two if the descending part of the light curve was also covered).

\begin{figure}
\includegraphics[width=0.93\textwidth]{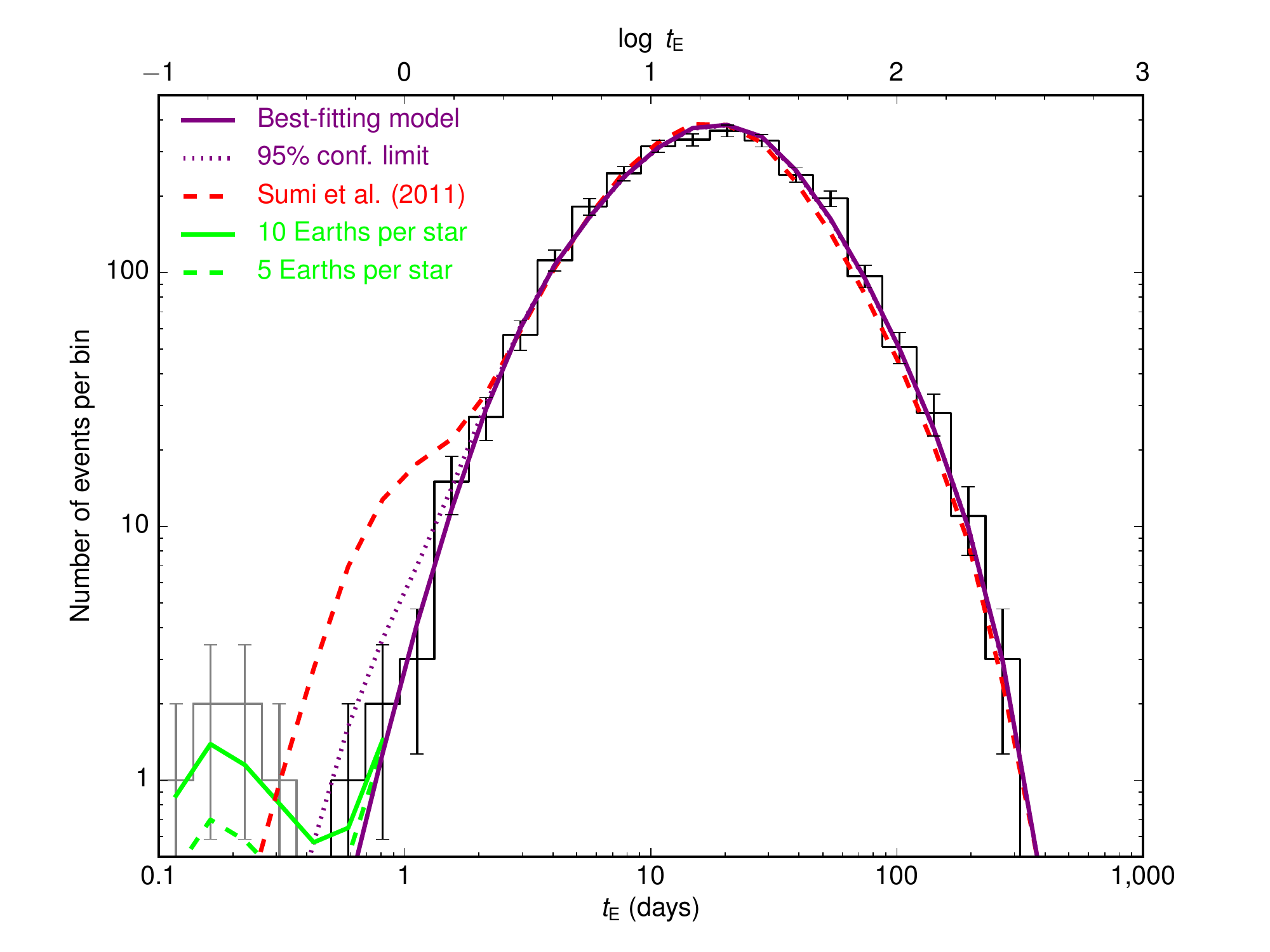}
\caption{\textbf{Observed distribution of timescales of 2,617 high-quality microlensing events discovered by OGLE in 2010--2015.} The purple line is the best-fitting model. The dotted line constrains the 95\% confidence limit on the number of wide-orbit or unbound Jupiter-mass planets of 0.25 planets per star. The dashed red line is the best-fitting model from ref. \cite{sumi} predicting almost two Jupiter-mass free-floating planets per star. According to that model we should find 64 events with $0.3<\tE<1.8$ d, but only 21 were observed (the discrepancy is even larger for events with $0.3<\tE<1.3$ d, where 6 events were found out of 42 expected). We detected six possible ultrashort-timescale events ($\tE<0.5$ d), which may be due to Earth-mass free-floating planets (grey histogram). Solid (dotted) green lines mark the expected microlensing signal assuming $5\ M_{\oplus}$ planets five (ten) times more frequent than stars. Error bars are the $1\sigma$ Poisson uncertainties on the counts of the number of events observed in a given $\tE$ bin.}
\label{fig:observed}
\end{figure}

Using our detection efficiency simulations (see below), we found that the event timescales cannot be reliably measured for faint, highly-blended events (with blending parameter $f_{\rm s} = F_{\rm s}/(F_{\rm s}+F_{\rm b})<0.1$, that is, less than 10\% of the baseline flux comes from the source), which was predicted theoretically\cite{wozniak}. Therefore, to ensure that our final results were robust, highly-blended events were not included in our sample of high-quality events. Thus, regardless of the timescale, there can be no systematic shift between measured and real timescales for simulated data. The final distribution of the event timescales is shown in Fig. \ref{fig:observed}.

To calculate the detection efficiency we conducted extensive image-level simulations, in which artificial microlensing events were injected into real OGLE images using the point spread function derived from the neighbouring stars. In total, 8.6 millions of artificial events were simulated. Parameters $u_0$, $t_{0}$, and $\log{t_{\rm E}}$ were drawn from uniform distributions, but sources were randomly drawn from the luminosity function of each subfield. For simulated events we applied exactly the same selection criteria as those applied to the observed sample of events. Detection efficiency curves for all analyzed fields are shown in Extended Data Fig. \ref{fig:eff}. 

\begin{figure}
\includegraphics[width=0.93\textwidth]{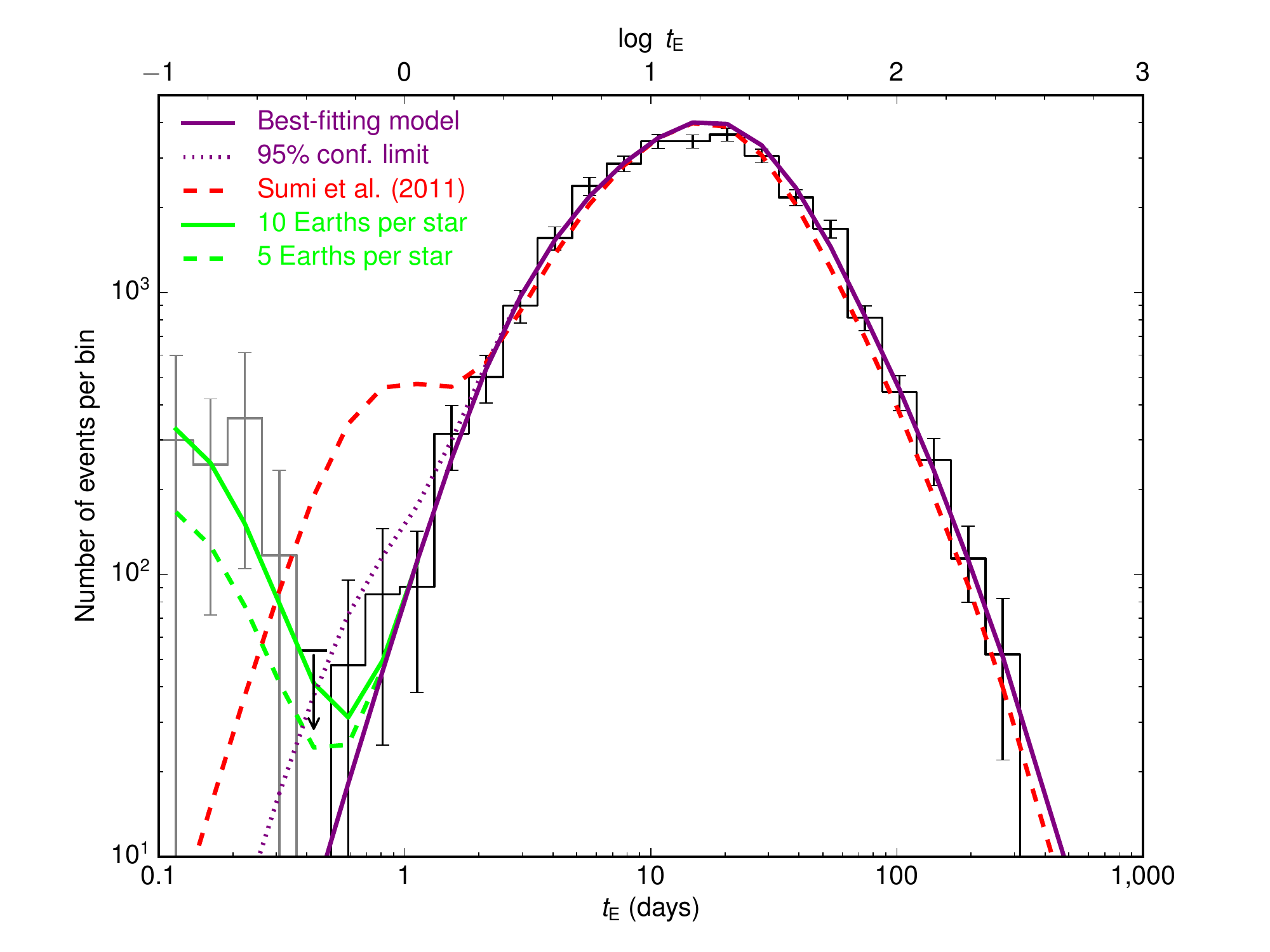}
\caption{\textbf{Distribution of event timescales corrected for the detection efficiency.} This distribution, at short timescales, can be well approximated as a power-law with a slope of $+3$, consistent with theoretical expectations\cite{mao1996}. There remains a small possible excess of events with timescales $0.5<\tE<1$ d. If they were caused by the Jupiter-mass lenses, the best-fitting models predict their frequency of 0.05 Jupiter-mass planets per star with a 95\% confidence limit of 0.25 planets per star (dotted purple line). All symbols are the same as in Fig. \ref{fig:observed}. Error bars are the $1\sigma$ Poisson uncertainties on the counts of the number of events observed in a given $\tE$ bin.}
\label{fig:best}
\end{figure}

The detection-efficiency-corrected histogram of event timescales is presented in Fig. \ref{fig:best} and, clearly, does not show the excess of events with timescales $\tE\approx 1-2$ d, claimed in ref. \cite{sumi}. The difference (at a confidence level of $2.5-3\sigma$) can be explained  in part by the relatively small number of events found in the earlier analysis\cite{sumi}. In addition to the 2,617 events analysed in this work, we detected over twenty short-duration events that showed clear signatures of binarity\cite{bennet} and did not pass our strict selection criteria for the fit quality. Owing to lower photometric precision, such events may have been mistaken for single short-timescale events. It is also possible that event timescales measured in the previous work suffer from systematic effects (differential refraction, unphysical treatment of negative blending). Thanks to better image quality (smaller pixel scale, better seeing) and a narrower filter, our photometry is less prone to such systematic effects.

We modelled the observed distribution of event timescales by maximizing the likelihood function $\mathcal{L}=\prod_{i} p(t_{{\rm E},i})$, where $p(t_{{\rm E}})=p_{\rm model}(t_{{\rm E}})\varepsilon(t_{{\rm E}})$ is the normalized predicted timescale distribution (corrected for the detection efficiency $\varepsilon(t_{{\rm E}})$)\cite{sumi,calchi}. We adopted a standard Galactic model\cite{han1995,han2003} of the distribution and kinematics of stars and tested several mass functions. 
In our best-fitting model, the initial mass function (IMF) can be approximated as a broken power law with slopes $-0.8$ in the brown dwarf regime ($0.01 < M < 0.08\ M_{\odot}$), $-1.3$ for low-mass stars ($0.08 < M < 0.5\ M_{\odot}$), and $-2.0$ for $M>0.5\ M_{\odot}$. We assumed that all stars more massive than $1\ M_{\odot}$ evolved into white dwarfs, neutron stars and black holes, depending on their initial mass, and we assumed the binary fraction $f_{\rm bin}=0.4$. The model is marked with a purple line in Figs \ref{fig:observed} and \ref{fig:best}.

Our best-fitting model describes the observed timescale distribution well, but we found there remains a small possible excess of events with timescales $0.5 < \tE < 1$ d. If we assume that they can be attributed to Jupiter-mass lenses\cite{sumi} ($M_{\rm lens}=10^{-3}\ M_{\odot}$), the maximum-likelihood models predict their frequency of 0.05 per main-sequence star with a 68\% confidence interval of $[0,0.12]$ planets per star. The 95\% confidence limit is 0.25 Jupiter-mass planets per star. These results agree with upper limits on the frequency of Jupiter-mass planets inferred from direct imaging surveys\cite{lafreniere2007,bowler2015}, which suggests that almost the entire possible excess of events with timescales $0.5 < \tE < 1$ d can be attributed to planets on wide orbits\cite{clanton}.

\begin{figure}
\includegraphics[width=0.8\textwidth]{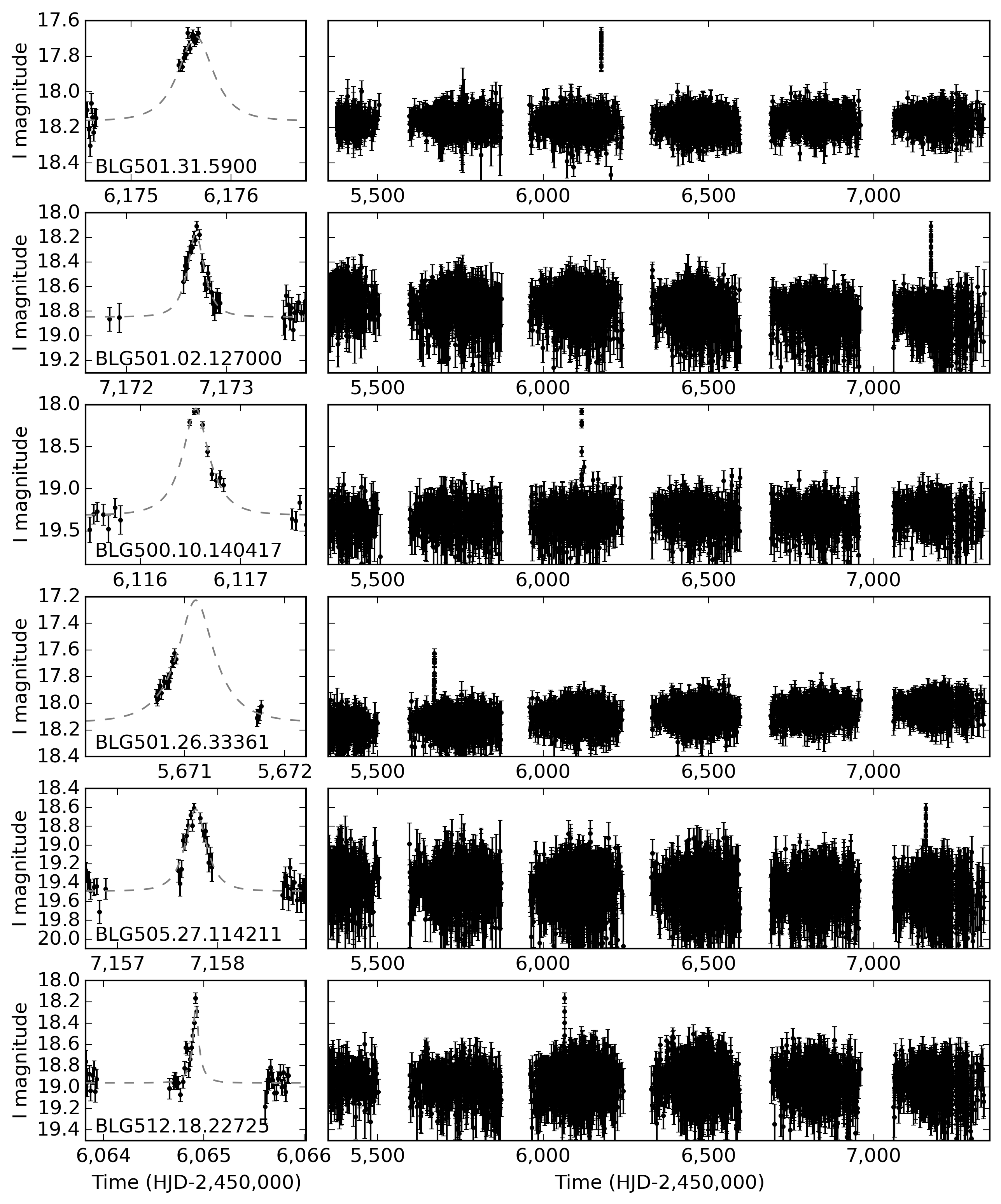}
\caption{\textbf{Light curves of ultrashort microlensing event candidates.} The left panels show a close-up of the light curve at the event and the right panels show 5.5-year long light curves from OGLE-IV. Some of those events have been observed by OGLE for 20 years with no trace of other variability, but we nevertheless cannot exclude the possibility that some of them may be flaring stars. The shortest-timescale events are not well covered by observations and it is difficult, if not impossible, to either prove or disprove their nature as free-floating planets. The detection efficiency at these timescales is very low, meaning that a very few detections imply the existence of a large population of Earth-mass free-floating or wide-orbit planets. Future space-based missions, like WFIRST and Euclid, will enable the exploration of these short events in more detail. Error bars represent $1\sigma$ uncertainties. HJD, Heliocentric Julian date.} 
\label{fig:short1}
\end{figure}

The timescales of six events passing our criteria for high-quality events are shorter than 0.5 day and these events last less that one night (Fig. \ref{fig:short1}). We carefully checked CCD images by eye to ensure that these brightenings are real, which rules out problems such as photometry artefacts or asteroids. We also analysed historical light curves for these events; four of the six have been observed by the OGLE survey for 20 years and we did not find any evidence for other outbursts in archival data. Nevertheless, because these events were so short and the light curves were not fully covered, we cannot rule out the possibility that some of them might be flaring stars (especially BLG512.18.22725 and BLG500.10.140417). 

The best-fitting microlensing models of six short events constrain their Einstein timescales in the range $0.1<\tE<0.4$ d (Extended Data Table \ref{tab:ultra}). Such short events should be caused by Earth- and super-Earth-mass objects, provided that they have kinematics that are similar to the brown dwarf, stellar and remnant lenses. They might be gravitationally unbound to any star or located at wide orbits (at least several astronomical units from the host star), given no signs of binarity in their light curves. Because the number of ultrashort events is very small and their nature is uncertain,  we do not attempt to model their mass function. However, a mere detection of such ultra-hort events means that Earth-mass lenses must be very common. If we assume that $5\ M_{\oplus}$-mass planets are five times more common than main-sequence stars, the expected number of ultrashort microlensing events is 2.2. For a more realistic mass function in which Earth-mass planets\cite{mao} are five times more common than main-sequence stars, the expected number of detections is 25\% smaller. 

According to planet formation theories, most Earth- and super-Earth-mass planets should form at relatively small orbital separations ($< 10$ au)\cite{ida}. The most likely sources of wide-orbit and free-floating Earth-mass planets are dynamical interactions in young multi-planet systems\cite{mao,pfyffer,barclay}. Other mechanisms (including ejections from multiple-star systems, stellar fly-bys, interactions in stellar clusters, and post-main-sequence evolution of the host star(s)) have also been proposed\cite{veras}. Although these processes are unlikely to produce a sizable population of Jupiter-mass free-floating planets, Earth-mass planets can be scattered and ejected much more efficiently.

Thanks to the superb photometry quality and the possibility of continuous observations during approximately $100$-day-long windows, future space-based missions, such as WFIRST\cite{spergel} and Euclid\cite{penny}, will have the potential to explore the population of free-floating Earth-mass planets in more detail.

\begin{addendum}
 \item[Acknowledgements] We thank M. Kubiak and G. Pietrzy{\'n}ski, former members of the OGLE team, for their contribution to the collection of the OGLE photometric data over the past years. The OGLE project has received funding from the National Science Center, Poland through grant MAESTRO 2014/14/A/ST9/00121 to A.U. 
 \item[Author Contributions] P.M. analysed and interpreted the data, and prepared the manuscript. A.U. initiated the project, reduced the data, and conducted detection efficiency simulations. All authors collected the OGLE photometric observations, reviewed, discussed and commented on the present results and on the manuscript.
 \item[Author Information] Reprints and permissions information is available at www.nature.com/reprints. The authors declare no competing financial interests. Readers are welcome to comment on the online version of the paper. Correspondence and requests for materials should be addressed to P.M. (pmroz@astrouw.edu.pl).
\end{addendum}

\newpage

\newpage

\begin{methods}

\subsection{Data.}

All data presented in this paper were collected as part of the OGLE-IV sky survey\cite{udalski} during the years 2010-2015. The survey uses the 1.3-m Warsaw Telescope, located at Las Campanas Observatory, Chile. The observatory is operated by the Carnegie Institution for Science. The telescope is equipped with a mosaic, 32-chip CCD camera covering a field of view of 1.4 sqaure degrees with a pixel scale of $0.26''$ per pixel. All objects analysed are located within nine OGLE fields, observed with a cadence of either 20 min (BLG501, BLG505, and BLG512) or 60 min (BLG500, BLG504, BLG506, BLG511, BLG534, and BLG611), covering in total 12.6 square degrees. We analyzed data collected between 2010 June 29 and 2015 November 8, that is, five and a half Galactic bulge observing seasons. Light curves consist of 4,500 -- 12,000 data points, depending on the field, which gives a total of 380 billion photometric measurements. All analyzed data were taken through the $I$-band filter. Basic information about the fields analyzed is presented in Extended Data Table \ref{tab:fields}.

OGLE photometric pipeline is based on the difference image analysis method (DIA)\cite{alard, wozniakdia}. For each field, a reference image is constructed by stacking several highest-quality and seeing frames. This reference image is then subtracted from incoming frames and the photometry is performed on subtracted images. Variable and transient objects that are detected on subtracted images are stored in two databases. The ``standard'' database consists of all stellar-like objects detected on the reference frame, whereas ``new'' objects (those that do not correlate with any identified stars) are stored separately; see the description of the OGLE photometric pipeline\cite{wozniakdia,udalski2003}.

\subsection{Event selection.}

We analysed 50 million light curves, from all the objects from the ``standard'' database.
We began our analysis by correcting photometric uncertainties\cite{skowron2016} and transforming magnitudes into flux. It is known that uncertainties returned by the DIA are underestimated and ref. \cite{skowron2016} provides an algorithm for their correction, so that these uncertainties now reflect the real observational scatter in the data. The selection criteria for high-quality microlensing events are summarized in Extended Data Table \ref{tab:crit}.

\textbf{Cut 1.} We placed a 360-day moving window on each light curve and measured the baseline flux $F_{\rm base}$ and its dispersion $\sigma_{\rm base}$ using data points outside the window (after rejecting $5\sigma$ outliers such as cosmic ray hits). We required $\chi^2_{\rm out}/{\rm d.o.f.} \leq 2.0$, where d.o.f. are degrees of freedom, outside the window, so we could reject most of the variable stars. Some genuine microlensing events with variable baseline or those longer than one year may have not passed this criterion.
We defined a \textit{bump} as a brightening with at least three consecutive points at least $3\sigma_{\rm base}$ above the baseline flux. For each bump we calculated $\chi_{3+}=\sum_i(F_i-F_{\rm base})/\sigma_i$ ($i$ is the index within a bump) and $n_{\rm DIA}$, the number of detections on subtracted images. We required $\chi_{3+}\geq 32$ and $n_{\rm DIA} \geq 3$ to pass this cut. (We note that with the current data we were able to set a lower threshold than in ref. \cite{sumi}, who used $\chi_{3+}\geq 80$). The introduction of the cut on $n_{\rm DIA}$ allowed us to eliminate contamination from asteroids, photometry artefacts, and ``ghost'' microlensing events, which are stars affected by real variability of neighbouring stars\cite{wyrzykowski2015}.

\textbf{Cut 2.} Cut 1 criteria were insufficient to remove all artefacts. For example, reflections within the telescope might cause spurious, short brightenings of neighboring stars correlated in time. Reflections were especially troublesome near the edges of CCD detectors \#1, \#7, \#8, \#16, \#17, \#25, \#26, and \#32 of the OGLE-IV mosaic camera, located at the edges of the telescope field of view\cite{udalski}. To quantify the concurrence of bumps, we defined the \textit{similarity} of two bumps as $s = N_1/N_2$, where $N_1$ is the number of individual frames when \textit{both} bumps were detected on subtracted images and $N_2$ is the number of frames when \textit{at least} one bump was detected. We calculated similarities for all possible pairs of bumps shorter than five days and then rejected objects with $s \geq 0.4$. This threshold value was chosen after visual inspection of light curves and images of possible short events. It allowed us to reject over 95\% of artefacts, while removing none of the genuine microlensing events from the sample.

A number of stars that passed cut 1 criteria were OGLE small amplitude red giants (OSARGs)\cite{eyer} which are red-giant variable stars showing low-amplitude ($<0.13$ mag in the $I$ band) pulsations with (frequently multiple) periods in the range $10 < P < 100$ d. Some pulsation cycles in OSARGs might have slightly higher amplitudes so they were detected by our algorithm as potential microlensing events. We therefore rejected all objects with a bump amplitude $A \leq 0.1$ mag, so only a few genuine microlensing events were discarded in this step. The remaining OSARGs were easily rejected in the next step, because the microlensing light curve fit yielded nonphysical parameters.

Finally, we rejected all objects with more than one bump in the light curve. These were mostly dwarf novae and some remaining photometry artefacts. Twenty-nine genuine microlensing events were also rejected, most of them binary source or binary lens events, and some microlensing events with variable baseline.

\textbf{Cut 3.} For the remaining 11,989 event candidates, we fitted the microlensing point-source point-lens model. The lensing model has three parameters: the time $t_0$ and projected separation $u_0$ (in Einstein radius units) between the lens and the source during the closest approach, and the Einstein radius crossing time $\tE$. The source flux $F_{\rm s}$ and the blend flux $F_{\rm b}$ were found analytically using the least-squares method. We also calculated the four-parameter fits, where the blend flux was set to zero, $F_{\rm b}=0$.
We performed the initial fit using the simplex algorithm using the data from a 360-day window centred on the event and later refined the parameters using all available data. 

We calculated a number of goodness-of-fit statistics. $\chi^2_{\rm fit}$ for the entire dataset, $\chi^2_{\rm fit,\tE}$ for $|t-t_0|<\tE$, $\chi^2_{\rm fit,2\tE}$ for $|t-t_0|<2\tE$, and $\chi^2_{{\rm fit,}k }$ for $|t-t_0|<k$ (where $k=1$ or $k=5$ days). We removed $4\sigma$ outliers provided that adjacent datapoints are within $1\sigma$ from the best-fitting model and $|t_{i\pm 1}-t_{i}|<1$ day. We required $\chi^2 /{\rm d.o.f.} \leq 2.0$, which removes the majority of non-standard microlensing events (finite source, parallax, binary) in addition to non-microlensing events. We allowed for some amount of negative blending, that is, the blend flux $F_{\rm b} > -F_0$ was allowed, where $F_0=0.251$ is the flux corresponding to an $19.5$-mag star (here $F=1$ corresponds to an 18-mag star). If $F_{\rm b} < -F_0$ and the four-parameter model was marginally worse ($\Delta\chi^2 < 4$) than the five-parameter model, we chose the four-parameter model. Usually, a high negative blending indicates that the single lensing model has been fitted to a non-microlensing event (like a dwarf nova outburst, OSARG, or stellar flare). However, a small amount of negative blending does not necessarily mean that the model is unphysical. The background (mainly unresolved main-sequence stars) in crowded fields of the Galactic bulge is not uniform and if the source happens to be located in a lower-density region, the blend flux might be negative. The issue of negative blending is discussed by refs \cite{park,jiang,smith}. We checked that our prior on the negative blending has no impact on the final event timescale distribution (which remains the same after choosing $F_0 = 0.1$, that is, the flux corresponding to a $20.5$-mag star).

We also required at least $n_{\rm r} \geq 2$ datapoints on the rising part of the light curve ($t_0-t_{\rm E} < t < t_0$) and at least $n_{\rm d} \geq 2$ datapoints on the descending branch ($t_0< t < t_0+t_{\rm E}$). If $n_{\rm d} < 2$, we required $n_{\rm r} \geq 4$. These cuts allowed us to eliminate contamination from flaring stars, which can rise very steeply\cite{hawley} (within minutes), but fade slowly (on a timescale of hours). If the rising part of the light curve is not sufficiently sampled, a flare might be mistaken for a very short microlensing event.

Our image-level simulations (see below) showed that we were unable to robustly measure the true timescale of an event if the event is faint and the blending is high ($f_{\rm s}<0.1$, that is, less than 10\% of baseline flux comes from the source). Therefore, to ensure that the final results are sound we did not include events with blending parameter $f_{\rm s}<0.1$. The inclusion of highly-blended events had little effect on the final results, although we found an increased number of long-timescale events ($t_{\rm E} > 100$ d).

The purity of our sample is almost 100\%. Over 90\% of microlensing events detected in the real-time by the OGLE Early Warning System\cite{udalski2003} passed our ``cut 2'' criteria. We detected additional 20--30\% events (depending on the field) compared to Early Warning System detections. The final distribution of timescales of detected microlensing events is shown in Fig. \ref{fig:observed}. Extended Data Table~\ref{tab:obs} presents the number of events detected in individual fields and timescale bins.

\subsection{Detection Efficiency.}

To calculate the event detection efficiency, we carried out extensive image-level simulations in which we injected artificial microlensing events into real OGLE frames using the PSF derived from neighboring stars. In each iteration we simulated 5,000 events per CCD detector, so the star density did not increase much (by 5--10\%). We carried out six iterations for each field, so in total 8.6 million of events were simulated in all fields.

Parameters $t_0$ and $u_0$ were drawn from uniform distributions: $0.0 \leq u_0 < 1.5$ and $2455377 \leq t_0 < 2457388$. Einstein timescales were drawn from a log-uniform distribution $-1.0 \leq \log\tE < 2.5$. Sources were taken from the range $14 \leq I_{\rm s} < 22$ mag from the luminosity function of each subfield, which was created as follows. We constructed a very deep luminosity function for the subfield BLG513.12, which was observed both by the OGLE-IV survey and the Hubble Space Telescope\cite{holtz1998}. The OGLE-IV luminosity function and the Hubble Space Telescope luminosity function overlap in the range $16 < I < 18$ mag (Extended Data Fig. \ref{fig:lf}). This deep luminosity function was used as a template to generate artificial microlensing events in other fields, after shifting it so that the centroid of the red clump matched the observed centroid. We therefore took into account variable bulge geometry and reddening. If there was evidence for differential reddening, we divided subfields into smaller parts. There were a few subfields (7\% of the total analysed area) where we were not able to detect the red clump owing to extremely high extinction; these were omitted from the final calculations (we detected only a negligible number of 48 microlensing events in these fields).

For the simulated events we applied exactly the same selection criteria as for the real events (Extended Data Table \ref{tab:crit}). The detection efficiency curves for all analysed fields are shown in Extended Data Fig. \ref{fig:eff} and listed in Extended Data Tab. \ref{tab:eff}. We note that detection efficiency for events with $\tE=2$ d is very high, up to 53\% of the maximum efficiency for field BLG512. Efficiencies for fields observed with 20-min and 60-min cadence are very similar, except for the shortest events with $t_{\rm E}< 0.5$ day. In general, we found that detection efficiencies are most sensitive to crowding and interstellar reddening toward the given field (fields with higher reddening and higher crowding have lower efficiencies). We note that events were simulated using a standard point-lens point-source model. Higher-order effects, like the parallax (causing deviations in the light curve induced by the Earth's motion\cite{gould1992}), were not included and so detection efficiencies for long events ($\tE \geq 100$ d) may be slightly overestimated. Similarly, we did not include the finite source effect, which may reduce our detection efficiency for the shortest events ($\tE \sim 0.1$ d), when the Einstein ring size becomes similar to the source star radius\cite{bennett}.

\subsection{Parameter recovery.}

We also used our simulations to ensure that there is no systematic difference between measured and real timescales. In Extended Data Fig. \ref{fig:in_out} we plot timescales for simulated events passing all criteria from Extended Data Table \ref{tab:crit}. We found there is no systematic bias in measured timescales, unless events were faint and highly blended. This effect was predicted by ref. \cite{wozniak}, where it was found theoretically that in such cases the event timescale, impact parameter and blending parameter may be severely correlated, because information on the event timescale comes mostly from wings of the light curve that can be more easily affected by the photometric noise. In Extended Data Fig. \ref{fig:blending}a we show the ratio between measured and ``real'' (simulated) timescale $t_{\rm E,out}/t_{\rm E,in}$ versus the blending parameter $f_{\rm s}=F_{\rm s}/(F_{\rm s}+F_{\rm b})$. It is clear that timescales of highly blended and faint events are not well measured and systematically overestimated. A similar effect was also noticed in the earlier work\cite{sumi}, where it was found that $t_{\rm E,in}$ was systematically about $5\%$ smaller than $t_{\rm E,out}$ regardless of $t_{\rm E}$. Strong correlations between blending, impact parameter, and event timescale may also lead to the incorrect determination of parameters. For example, one of short events reported by ref. \cite{sumi}, MOA-ip-1, has incorrectly measured timescale. The best-fitting model with $\tE = 8.2^{+8.1}_{-3.6}$ d is better by $\Delta\chi^2 = 9$ than the model presented in the original paper ($\tE = 0.73 \pm 0.08$ d).

To be conservative, we decided not to include highly-blended events ($f_{\rm s} < 0.1$) in our final sample of high-quality events. Thanks to this selection cut, there is almost no bias in the measured timescales (see Extended Data Figs. \ref{fig:in_out} and \ref{fig:blending}b). 

\subsection{Modeling Timescale Distribution.}

The actual timescale distribution depends on the distribution and kinematics of lenses and sources as well as the underlying mass function\cite{kiraga1994,han1996,mao1996}. The timescale distribution can be computed from a multi-dimensional integral\cite{han1996,bissantz2004}:
\begin{align*}
f(\tE) \propto& \int \rho(D_S) \rho(D_L) R_{\rm E} (D_L, D_S, M) \Phi (M) \\
& \times v_{\rm rel} f(v_{\rm rel}) \delta(\tE-\frac{R_{\rm E}}{v_{\rm rel}}) dD_L dD_S dv_{\rm rel} dM,
\end{align*}
where $\rho(D)$ is the distribution of lenses and sources along the line-of-sight, $R_{\rm E}=\theta_{\rm E}D_{\rm L}$ the Einstein radius, $v_{\rm rel}$ is the lens-source relative velocity projected onto the plane of the sky, and $\Phi(M)$ is the mass function. We expect the timescale distribution to have power-law tails with slopes of $+3$ and $-3$ at short and long timescales, respectively\cite{mao1996,mao2005}.

To compare the measured distribution of Einstein timescales with models, we maximized the following log-likelihood function:
\begin{equation*}
\ln \mathcal{L}=\sum_{i} \ln p(t_{{\rm E},i}),
\end{equation*}
where $p(t_{{\rm E}})=p_{\rm model}(t_{{\rm E}})\varepsilon(t_{{\rm E}})$ is the normalized predicted timescale distribution, which serves as our likelihood function. Here $p_{\rm model}(t_{{\rm E}})$ is the timescale distribution from the Galactic model and $\varepsilon(t_{{\rm E}})$ is the detection efficiency in a given field. The summation was performed over all events. 
We adopted a standard Galactic model\cite{han1995,han2003}, which incorporates the boxy-shaped bulge model\cite{dwek1995} and the double exponential model of the Galactic disk\cite{zheng2001}.

\subsection{Mass function.}

A detailed modeling of the initial mass function (IMF) would require population synthesis calculations, in addition to more sophisticated Galactic models, which is beyond the scope of this work. However, we can obtain useful constrains on slopes of the IMF using a simple model. Here we followed the approach of ref. \cite{gould2000} and  we assumed that all stars with initial masses $1 < M/M_{\odot} \leq 8$ evolved into white dwarfs following the empirical initial-final mass relation for white dwarfs\cite{williams2009} $M_{\rm final} = 0.339 + 0.129\ M_{\rm init}$. Masses of neutron stars (with initial masses in the range $8 < M/M_{\odot} \leq 20$) peak around $1.33\ M_{\odot}$ with a 68\% confidence interval of $(1.21,1.43)\ M_{\odot}$ (ref. \cite{kiziltan}), while for black holes we assumed a Gaussian distribution at $7.8 \pm 1.2\ M_{\odot}$ (ref. \cite{ozel2010}).

We fitted the following initial mass function:
\begin{equation*}
\Phi (M) = \begin{cases}
    a_1 M^{-\alpha_{\rm bd}} & 0.01 M_{\odot} \leq M < 0.08 M_{\odot}\\
    a_2 M^{-\alpha_{\rm ms}} & 0.08 M_{\odot} \leq M < M_{\rm break} \\
    a_3 M^{-2.0} & M \geq M_{\rm break}\\
  \end{cases}.
\end{equation*}
We allowed $\alpha_{\rm bd}$ and $\alpha_{\rm ms}$ to vary, but we assumed a fixed IMF slope of $-2.0$ above $M>M_{\rm break}=0.5 M_{\sun}$ (ref. \cite{zoccali2000}), because our experiment was designed to analyze the low-mass end of the IMF. We also considered models with $M_{\rm break}=0.7 M_{\sun}$ and models with binary fraction $f_{\rm bin}\neq 0$, where we assumed a flat mass ratio distribution\cite{belczynski2008} $f(q)=1$ in a range $0\leq q \leq 1$.

We conducted modelling using events with $\tE > 0.5$ and $\tE > 2.0$ days and in both cases we obtained virtually identical results. Constraints on slopes of the IMF are shown in Extended Data Fig. \ref{fig:imf}. In general, we found that models with non-zero binary fraction describe the event timescale distribution better than models with $f_{\rm bin}= 0$. The standard IMF\cite{kroupa2001} with $f_{\rm bin}= 0$ does not describe the entire timescale distribution well, especially at long timescales $t_{\rm E}>50$ d, which has already been noted\cite{wegg2016}. This may indicate that the current Galactic model underpredicts the number of long-timescale events, or the mass function of remnants (especially black holes) is underestimated, or remnants have distinct kinematics from brown dwarf and stellar lenses. The discrepancy can be also explained, if we assume that some fraction of lenses ($f_{\rm bin}$) are binary systems. Our models with $f_{\rm bin}= 0.4$ are substantially better than with $f_{\rm bin}= 0.0$ (with improvement in log-likelihood $\Delta\chi^2=2.0(\ln\mathcal{L}_{\rm max,1}-\ln\mathcal{L}_{\rm max,2})=18.6$). For the best-fitting models $\alpha_{\rm bd}\approx 0.8$ and $\alpha_{\rm ms}\approx 1.3$ with $3\sigma$ confidence intervals: $0.2 < \alpha_{\rm bd} < 1.3$ and $1.1 < \alpha_{\rm ms}< 1.5$. This corresponds to $0.90\pm0.05$ ($1\sigma$) brown dwarfs per main-sequence star. Ref. \cite{sumi} obtained a slightly lower IMF slope in the brown dwarf regime of $\alpha_{\rm bd}=0.49^{+0.24}_{-0.27}$, but they used fixed $\alpha_{\rm ms}=1.3$ and $f_{\rm bin}= 0$ (their slope $\alpha_{\rm bd}$ is in fact consistent with our models from Extended Data Fig. \ref{fig:imf}a for fixed $\alpha_{\rm ms}$).

The IMF slope derived in the stellar regime is consistent with the ``canonical''\cite{kroupa2001} value of $-1.3$. Observations of brown dwarfs in open clusters and star-forming regions indicate $\alpha_{\rm bd}\approx 0.6-0.7$ (ref. \cite{oliveira} and references therein) and our models are consistent with those values. On the other hand, censuses of nearby field brown dwarfs tend to prefer lower slopes. Ref. \cite{allen2005} found a 60\% confidence interval of $\alpha_{\rm bd}\approx 0.3\pm0.6$ and other studies support $\alpha_{\rm bd}\sim 0$ (ref. \cite{oliveira}). However, mass function measurements for isolated field brown dwarfs are affected by difficulties in measuring their ages, distances, and masses.

\subsection{Planetary mass function.}

To explain the excess of short events, ref. \cite{sumi} modelled their event timescale distribution using a stellar IMF with $\alpha_{\rm bd} = 0.5$,  $\alpha_{\rm ms} = 1.3$, and $M_{\rm break}=0.7 M_{\sun}$ with additional planetary component, approximated as a delta function at $M=10^{-3}\ M_{\odot}$. That model is shown in Figs \ref{fig:observed} and \ref{fig:best} as dashed red line. According to that model we should find 64 events with $0.3<\tE<1.8$ d, but only 21 were observed (the discrepancy is even larger for events with $0.3<\tE<1.3$ d, where 6 events were found out of 42 expected). Moreover, model of ref. \cite{sumi} systematically underpredicts the number of long-timescale events (because of its very low sensitivity to long events, $\tE > 100$ d, they found only five events in this range).

Our best-fitting model describes the observed timescale distribution well, but there remains a small possible excess of events with timescales $0.5 < \tE < 1$ d (Figs \ref{fig:observed} and \ref{fig:best}). If we assume, following ref. \cite{sumi}, they are due to Jupiter-mass lenses ($M_{\rm lens}=10^{-3}\ M_{\odot}$), the best-fitting models predict their frequency of 0.05 Jupiter-mass planet per star with 68\% confidence interval of $[0,0.12]$ planets per star. The 95\% confidence limit is 0.25 Jupiter-mass planet per star. Our results agree with upper limits on the frequency of Jovian-mass planets inferred from direct imaging surveys \cite{lafreniere2007,quanz2012}. 
For example, a high-contrast adaptive imaging search\cite{bowler2015} for giant planets around nearby M-dwarf stars did not find any planets, providing very strong upper limits (at the 95\% confidence limit) of 10-16\% (depending on the model) for planets of between 1 and 13 Jupiter masses, at a distance of approximately $10-100$ AU. This suggests that almost the entire possible excess of events with timescales $0.5 < \tE < 1$ d can be attributed to planets on wide orbits.

\subsection{Code availability.}

We have opted not to make the event detection and simulation codes publicly available, because they were designed to work with internal photometric databases. The code for the modelling of the timescale distribution is available from the corresponding author upon reasonable request. 

\subsection{Data availability.}

The data that support the findings of this study are available from the corresponding author upon reasonable request.

\end{methods}

\newpage

\setcounter{figure}{0}
\renewcommand{\figurename}{Extended Data Figure}
\renewcommand{\tablename}{\textbf{Extended Data Table}}

\begin{figure}
\centering
\includegraphics[width=0.73\textwidth]{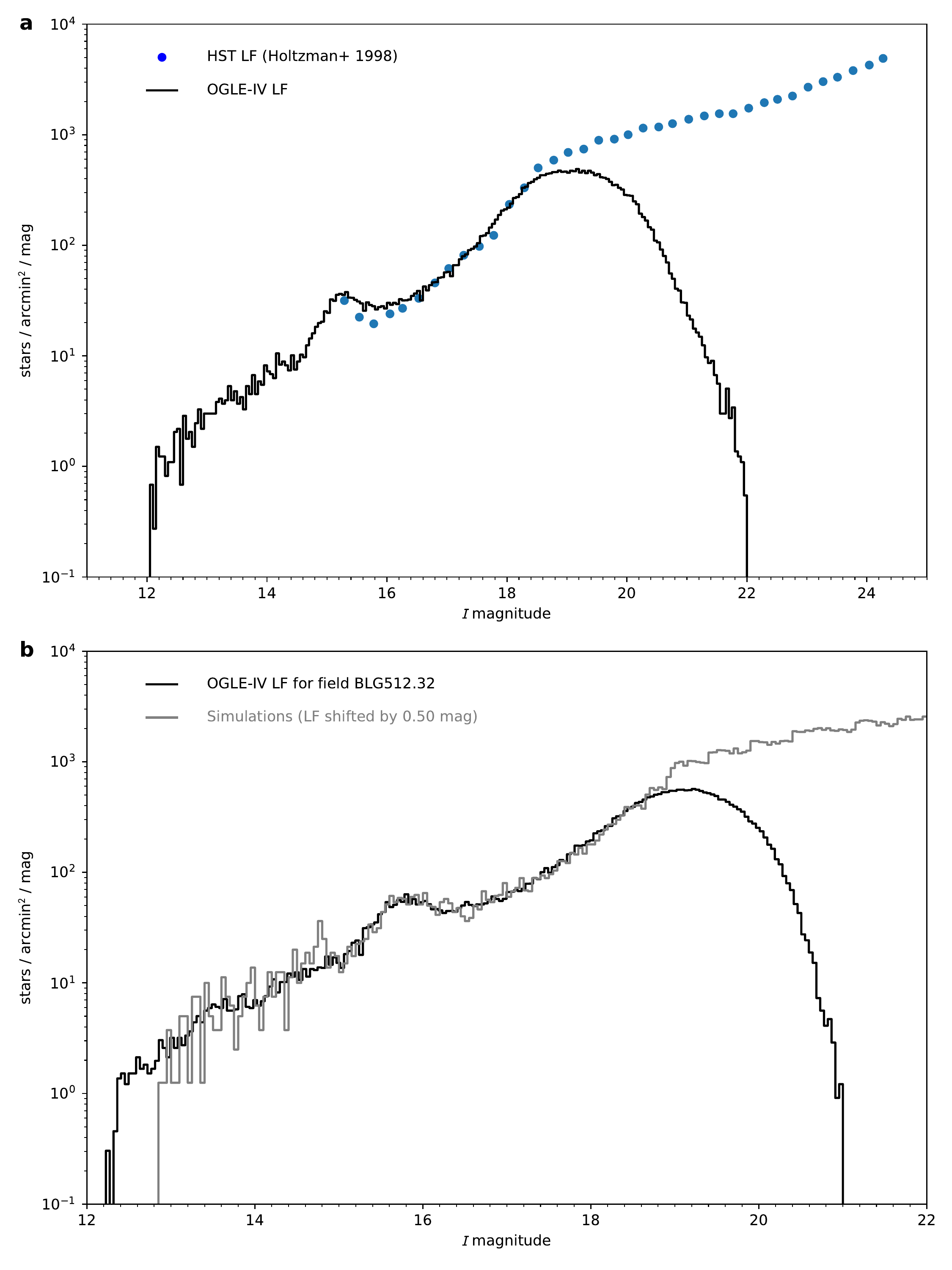}
\caption{\textbf{Galactic bulge luminosity function used for simulations.} \textbf{a,} Deep luminosity function (LF) for subfield BLG513.12, which was observed both by the OGLE-IV survey and the Hubble Space Telescope (HST)\cite{holtz1998}. Both LFs overlap in the range $16 < I < 18$ mag. This deep LF was used as a template to generate artificial microlensing events in analysed fields, after shifting to match the red clump's centroid in a given field. \textbf{b,} Comparison between the observed LF for subfield BLG512.32 and the simulated LF.}
\label{fig:lf}
\end{figure}

\begin{figure}
\includegraphics[width=0.96\textwidth]{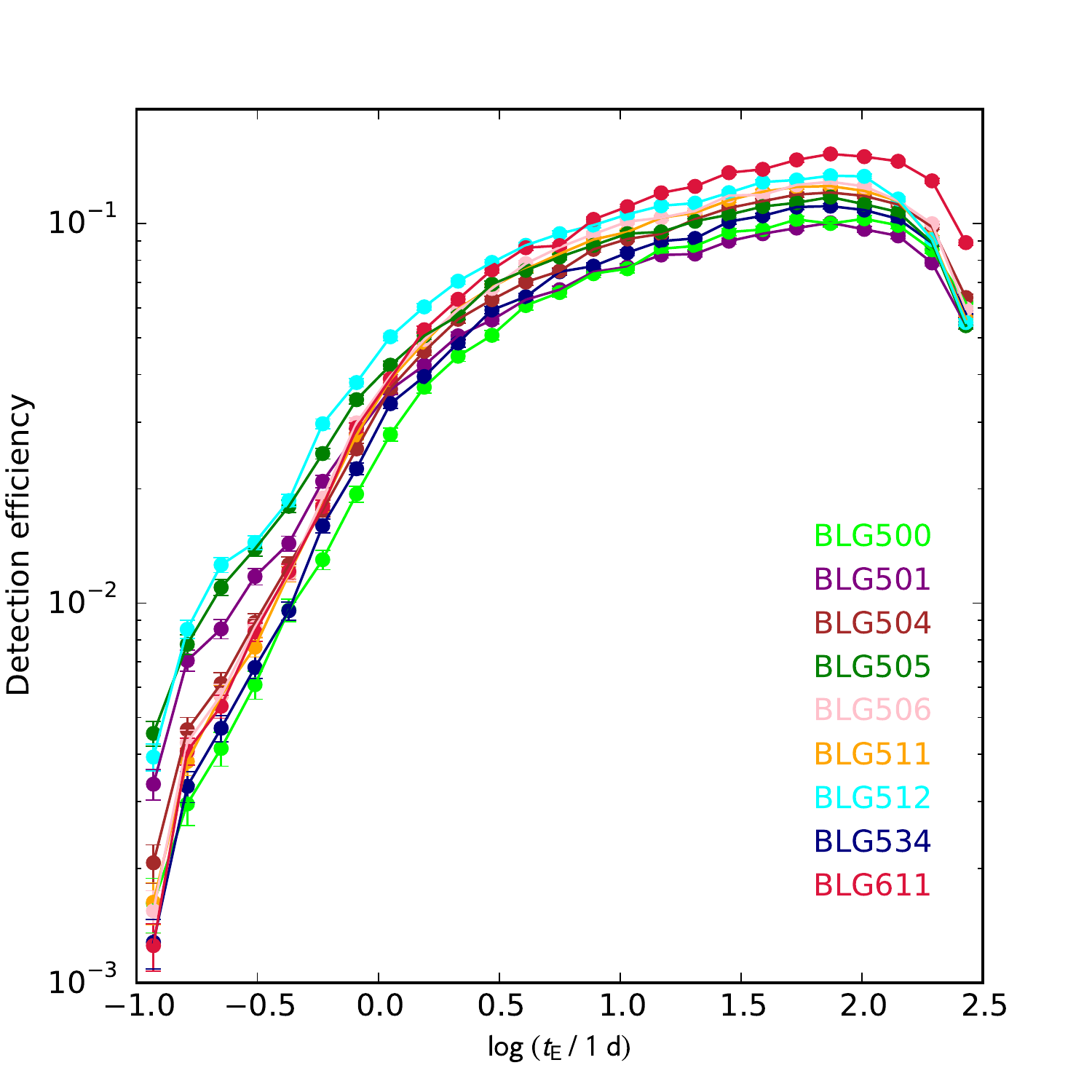}
\caption{\textbf{Detection efficiency curves.} Detection efficiencies as a function of the Einstein timescale $t_{\rm E}$ for all analysed fields (averages for all subfields in the given field). Fields BLG501, BLG505, and BLG512 were observed with a 20-min cadence, and the remaining fields with a 60-min cadence. Error bars are the $1\sigma$ Poisson uncertainties on the counts of the number of simulated events in a given $\tE$ bin.}
\label{fig:eff}
\end{figure}

\begin{figure}
\includegraphics[width=\textwidth]{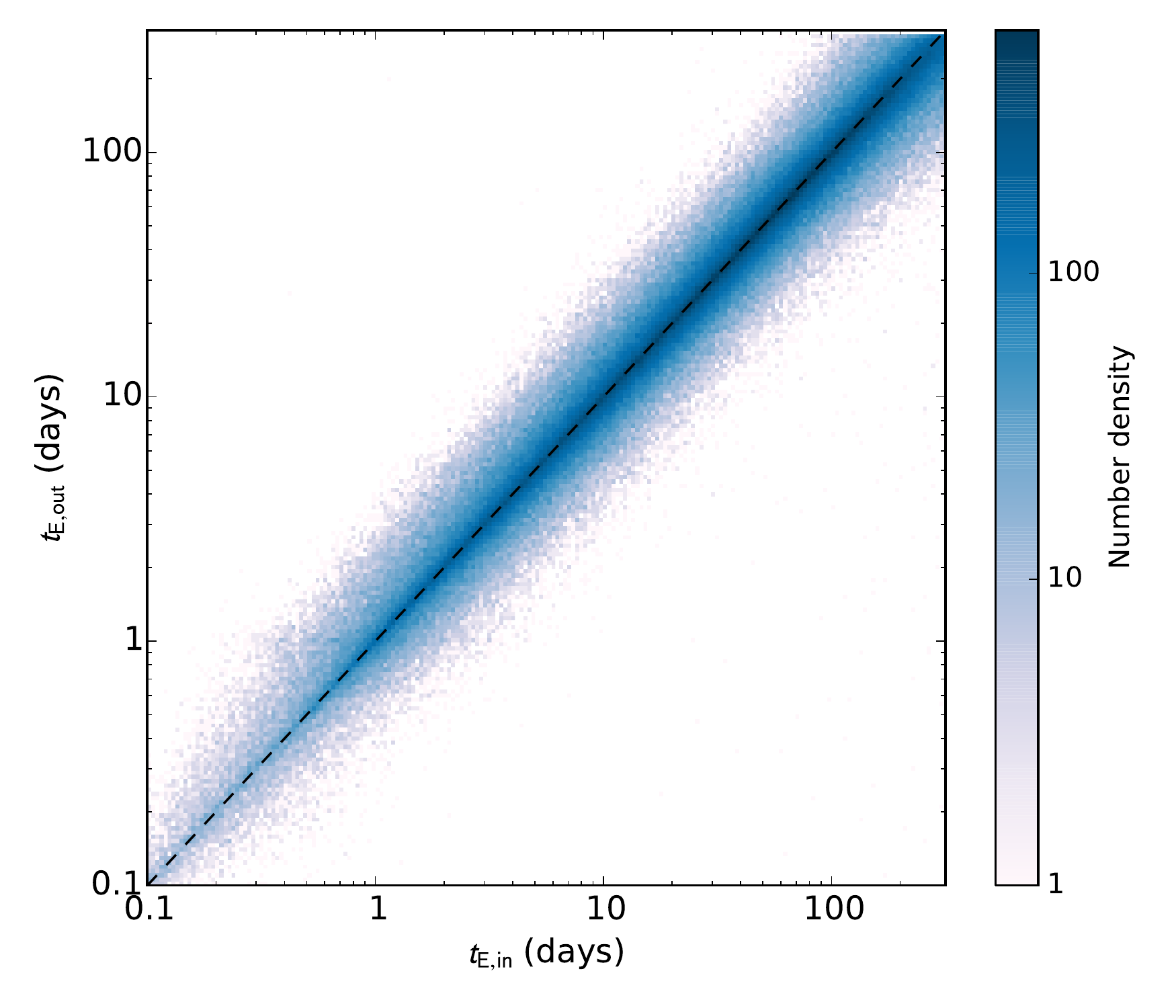}
\caption{\textbf{Comparison between measured Einstein timescales $t_{\rm E,out}$ and ``real'' (simulated) timescales $t_{\rm E,in}$ for simulated events.} Only events passing selection criteria from Extended Data Table \ref{tab:crit} (including the cut on the blending parameter $f_{\rm s}>0.1$) are shown. Note that the colour scale is logarithmic. There is no systematic offset between measured and real timescales.}
\label{fig:in_out}
\end{figure}

\begin{figure}
\centering
\includegraphics[width=\textwidth]{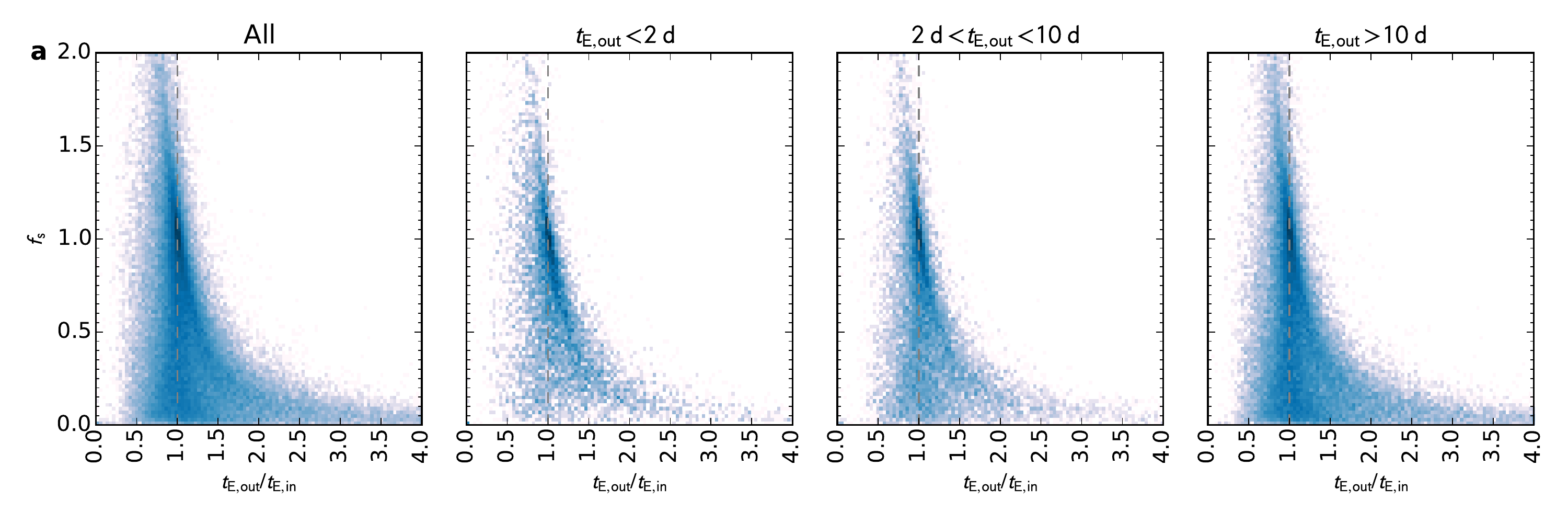}
\includegraphics[width=\textwidth]{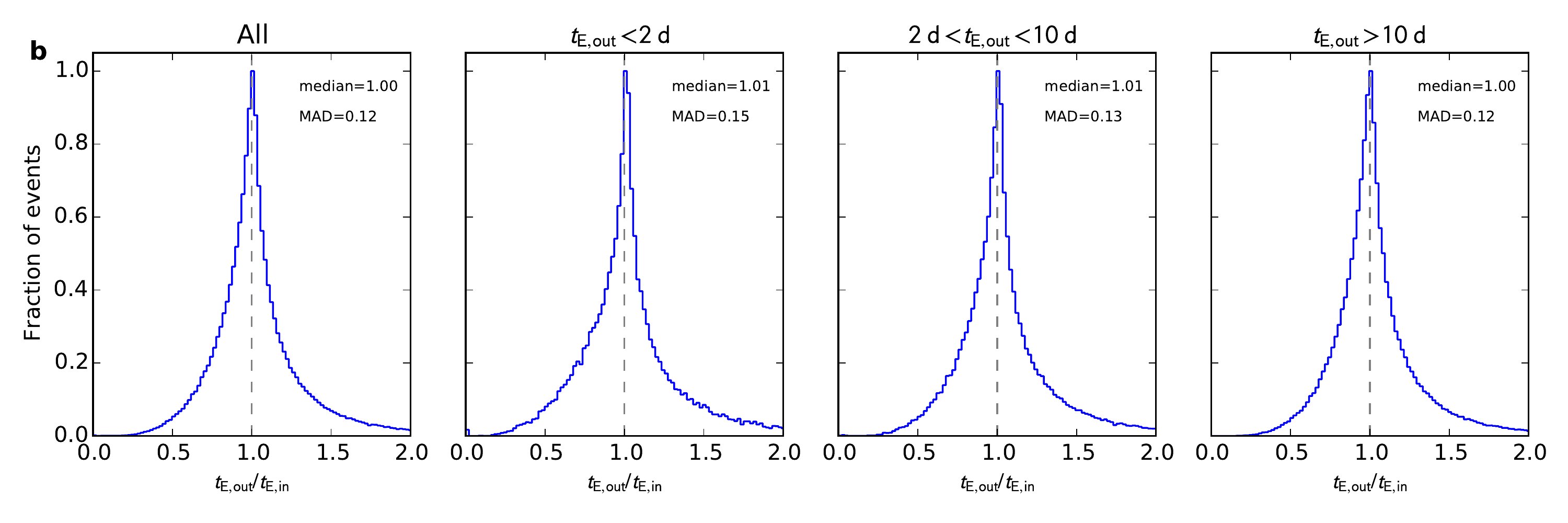}
\caption{\textbf{Comparison between measured and ``real'' (simulated) parameters.} \textbf{a,} Ratio between the measured Einstein timescale $t_{\rm E,out}$ and ``real'' (simulated) timescale $t_{\rm E,in}$ for simulated events versus the blending parameter $f_{\rm s}=F_{\rm s}/(F_{\rm s}+F_{\rm b})$. Timescales of faint and highly-blended ($f_{\rm s} < 0.1$) events are not well measured and are biased by a strong degeneration between Einstein timescale, blending and impact parameters. Timescales of events showing a high negative blending ($f_{\rm s}>1.5$) are systematically underestimated, but the bias is relatively small and such events comprise a negligible fraction of all events. \textbf{b,} Distributions of $t_{\rm E,out}/t_{\rm E,in}$ for simulated events passing selection criteria from Extended Data Table \ref{tab:crit} (including the cut on the blending parameter $f_{\rm s}>0.1$). Regardless of the timescale, there is no systematic bias between measured and real timescales within 1\%. For 90\% of simulated events $0.63 < t_{\rm E,out}/t_{\rm E,in} < 1.65$. The MAD is the median absolute deviation from the data's median.}
\label{fig:blending}
\end{figure}

\begin{figure}
\centering
\includegraphics[width=0.5\textwidth]{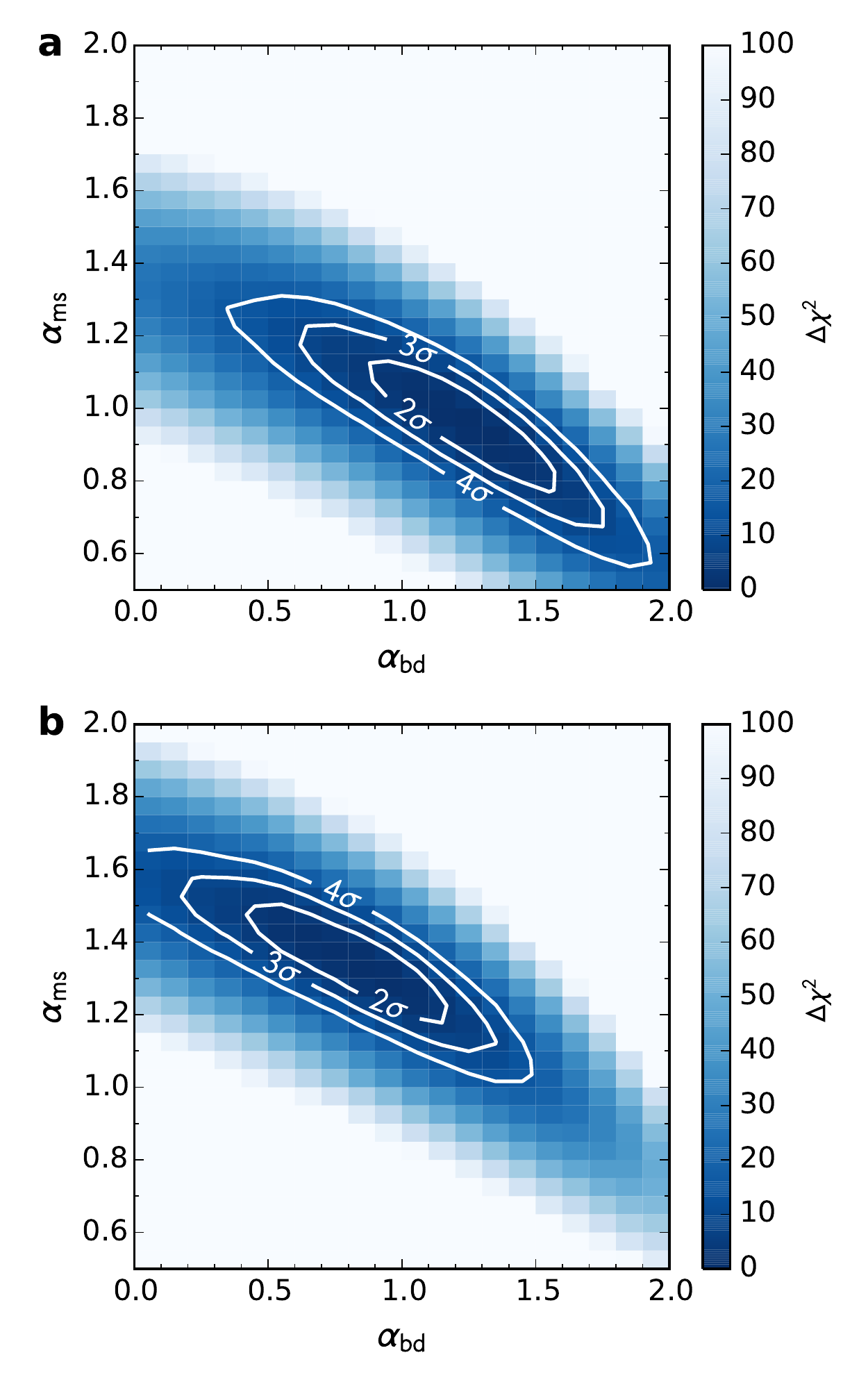}
\caption{\textbf{Constraints on IMF slopes:} \textbf{a,} Assuming that all lenses are single; \textbf{b,} assuming binary fraction $f_{\rm bin}= 0.4$.}
\label{fig:imf}
\end{figure}

\newpage

\begin{table}
\centering
\scriptsize
\begin{tabular}{lrrrrrrrr}
\hline
Star & RA & Decl. & $t_0$ & $\tE$ & $\tE$ $1\sigma$ conf.int. & $u_0$ & $I_{\rm s}$ & $f_{\rm s}$ \\
\hline
BLG501.31.5900   & 17:50:42.45 & -29:24:49.7 & 2456175.648 & 0.241 & [0.21,0.78] & 0.772 & 18.20 & 0.97 \\
BLG501.02.127000 & 17:53:13.44 & -30:18:59.6 & 2457172.692 & 0.146 & [0.12,0.26] & 0.517 & 19.13 & 0.77 \\
BLG500.10.140417 & 17:53:16.89 & -28:40:51.4 & 2456116.554 & 0.246 & [0.23,0.37] & 0.377 & 19.08 & 1.24 \\
BLG501.26.33361  & 17:54:17.54 & -29:18:17.0 & 2455671.124 & 0.320 & [0.29,0.79] & 0.471 & 18.04 & 1.11 \\
BLG505.27.114211 & 17:59:04.18 & -28:36:51.7 & 2457157.780 & 0.158 & [0.15,0.21] & 0.597 & 19.14 & 1.38 \\
BLG512.18.22725  & 18:05:25.00 & -28:28:23.9 & 2456064.921 & 0.128 & [0.08,0.19] & 0.138 & 20.95 & 0.16 \\
\hline
\end{tabular}
\caption{\textbf{Best-fitting parameters for ultrashort microlensing event candidates.} $I_{\rm s}$ is the source brightness and $f_{\rm s}=F_{\rm s}/(F_{\rm s}+F_{\rm b})$ is the blending parameter. The inclusion of the finite source effect does not improve $\chi^2$ much (typically $\Delta\chi^2=0.0-3.3$). Equatorial coordinates are given for the epoch J2000. We also show $1\sigma$ confidence intervals for $\tE$. RA, right ascension; Decl., declination.}
\label{tab:ultra}
\end{table}

\begin{table}
\centering
\scriptsize
\begin{tabular}{|l|l|l|r|r|r|r|}
\hline
Field & RA & Decl. & $l$ & $b$ & $N_{\rm stars}$ & $N_{\rm epochs}$\\ \hline
BLG500 &   17:51:60 &  -28:36:35 &   0.9999  &  -1.0293 & 4.0 & 4708 \\
BLG501 &   17:51:56 &  -29:50:00 & 359.9392  &  -1.6400 & 5.2 & 12117 \\
BLG504 &   17:57:33 &  -27:59:40 &   2.1491  &  -1.7747 & 5.8 & 6435\\
BLG505 &   17:57:34 &  -29:13:15 &   1.0870  &  -2.3890 & 6.9 & 12083\\
BLG506 &   17:57:31 &  -30:27:23 &   0.0103  &  -2.9974 & 5.3 & 4712\\
BLG511 &   18:03:02 &  -27:22:49 &   3.2835  &  -2.5219 & 5.5 & 4595\\
BLG512 &   18:03:04 &  -28:36:39 &   2.2154  &  -3.1355 & 6.9 & 10268\\
BLG534 &   17:51:51 &  -31:04:15 & 358.8644  &  -2.2547 & 4.2 & 4652\\
BLG611 &   17:35:33 &  -27:09:41 &   0.3282  &   2.8242 & 5.0 & 4526\\
\hline
\end{tabular}
\caption{\textbf{Basic information about analysed fields.} Equatorial coordinates are given for the epoch J2000. $N_{\rm stars}$ is the number of stars in millions and $N_{\rm epochs}$ is the number of observed frames during 2010--2015.}
\label{tab:fields}
\end{table}

\begin{table}
\scriptsize
\begin{tabular}{llr}
\hline
Criteria & Remarks & Number \\
\hline
$\chi^2_{\rm out}/{\rm dof} \leq 2.0$ & No variability outside the 360-day window centered on the event \\
$n_{\rm DIA} \geq 3$ & Centroid of the additional flux coincides with the source star centroid\\
$\chi_{3+}=\sum_i(F_i-F_{\rm base})/\sigma_i\geq 32$ & Significance of the bump & 43,158\\
\hline
$s<0.4$ & Rejecting photometry artifacts \\
$A>0.1$ mag & Rejecting low-amplitude variables \\
$n_{\rm bump}=1$ & Rejecting objects with multiple bumps & 11,989 \\
\hline
& Fit quality: & \\
$\chi^2_{\rm fit}/\rm{dof} \leq 2.0$ & $\chi^2$ for all data \\
$\chi^2_{\rm fit,\tE}/\rm{dof} \leq 2.0$ & $\chi^2$ for $|t-t_0|<\tE$ \\
$\chi^2_{\rm fit,2\tE}/\rm{dof} \leq 2.0$ & $\chi^2$ for $|t-t_0|<2\tE$ \\
$\chi^2_{\rm fit,1}/\rm{dof} \leq 2.0$ & $\chi^2$ for $|t-t_0|<1$ day \\
$\chi^2_{\rm fit,5}/\rm{dof} \leq 2.0$ & $\chi^2$ for $|t-t_0|<5$ days\\
$2455377\leq t_0\leq 2457388$ & Event peaked between 2010 June 29 and 2015 December 31 \\
$u_0 \leq 1$ & The minimum impact parameter \\
$I_{\rm s} \leq 22.0$ & The minimum $I$-band source magnitude \\
$n_{\rm r} \geq 2$ if $n_{\rm d} \geq 2$ & Rising and descending parts of the light curve should be sufficiently sampled\\
$n_{\rm r} \geq 4$ if $n_{\rm d} < 2$ & \\
$F_{\rm b} > -0.251$ &  The maximum negative blend flux, corresponding to $I=19.5$ mag star \\
$f_{\rm s} > 0.1$ &  Rejecting highly-blended events & 2617\\
\hline
\end{tabular}
\caption{\textbf{Selection criteria for high-quality microlensing events.}}
\label{tab:crit}
\end{table}

\begin{table}
\centering
\scriptsize
\begin{tabular}{|r|r|rrrrrrrrr|}
\hline
Bin & $\log\tE$ & BLG500 & BLG501 & BLG504 & BLG505 & BLG506 & BLG511 & BLG512 & BLG534 & BLG611 \\
\hline
1 & -0.93 & 0 & 1 & 0 & 0 & 0 & 0 & 0 & 0 & 0 \\
2 & -0.79 & 0 & 0 & 0 & 1 & 0 & 0 & 1 & 0 & 0 \\
3 & -0.65 & 1 & 1 & 0 & 0 & 0 & 0 & 0 & 0 & 0 \\
4 & -0.51 & 0 & 1 & 0 & 0 & 0 & 0 & 0 & 0 & 0 \\
5 & -0.37 & 0 & 0 & 0 & 0 & 0 & 0 & 0 & 0 & 0 \\
6 & -0.23 & 0 & 1 & 0 & 0 & 0 & 0 & 0 & 0 & 0 \\
7 & -0.09 & 1 & 0 & 0 & 0 & 1 & 0 & 0 & 0 & 0 \\
8 & 0.05 & 1 & 1 & 1 & 0 & 0 & 0 & 0 & 0 & 0 \\
9 & 0.19 & 0 & 4 & 0 & 2 & 0 & 4 & 1 & 1 & 3 \\
10 & 0.33 & 3 & 5 & 4 & 4 & 1 & 0 & 3 & 5 & 2 \\
11 & 0.47 & 4 & 9 & 7 & 8 & 5 & 8 & 3 & 8 & 5 \\
12 & 0.61 & 10 & 19 & 13 & 28 & 10 & 10 & 13 & 6 & 3 \\
13 & 0.75 & 17 & 40 & 17 & 39 & 19 & 11 & 13 & 17 & 9 \\
14 & 0.89 & 22 & 32 & 24 & 55 & 26 & 19 & 28 & 20 & 20 \\
15 & 1.03 & 25 & 39 & 30 & 78 & 34 & 22 & 40 & 22 & 25 \\
16 & 1.17 & 26 & 35 & 46 & 57 & 44 & 33 & 46 & 24 & 23 \\
17 & 1.31 & 29 & 62 & 38 & 62 & 39 & 30 & 40 & 24 & 38 \\
18 & 1.45 & 23 & 42 & 39 & 53 & 32 & 32 & 41 & 33 & 36 \\
19 & 1.59 & 15 & 39 & 27 & 40 & 32 & 24 & 25 & 20 & 21 \\
20 & 1.73 & 12 & 25 & 20 & 39 & 19 & 21 & 31 & 18 & 11 \\
21 & 1.87 & 7 & 13 & 11 & 20 & 10 & 10 & 12 & 6 & 8 \\
22 & 2.01 & 3 & 9 & 6 & 11 & 6 & 7 & 3 & 2 & 4 \\
23 & 2.15 & 5 & 2 & 3 & 2 & 7 & 1 & 4 & 2 & 2 \\
24 & 2.29 & 0 & 0 & 1 & 1 & 3 & 2 & 3 & 0 & 1 \\
25 & 2.43 & 0 & 0 & 0 & 1 & 2 & 0 & 0 & 0 & 0 \\
\hline
\end{tabular}
\caption{\textbf{Number of events detected in individual timescale bins.} There are 25 bins equally spaced in $\log\tE$ between $-1.0$ and $2.5$.}
\label{tab:obs}
\end{table}

\newpage

\begin{table}
\centering
\scriptsize
\begin{tabular}{|r|r|rrrrrrrrr|}
\hline
Bin & $\log\tE$ & BLG500 & BLG501 & BLG504 & BLG505 & BLG506 & BLG511 & BLG512 & BLG534 & BLG611 \\
\hline
1 & -0.93 & 0.0016 & 0.0033 & 0.0021 & 0.0045 & 0.0015 & 0.0016 & 0.0039 & 0.0013 & 0.0013 \\ 
2 & -0.79 & 0.0030 & 0.0071 & 0.0046 & 0.0078 & 0.0043 & 0.0038 & 0.0085 & 0.0033 & 0.0041 \\ 
3 & -0.65 & 0.0041 & 0.0086 & 0.0061 & 0.0110 & 0.0057 & 0.0057 & 0.0126 & 0.0047 & 0.0053 \\ 
4 & -0.51 & 0.0061 & 0.0118 & 0.0089 & 0.0139 & 0.0086 & 0.0077 & 0.0144 & 0.0068 & 0.0084 \\ 
5 & -0.37 & 0.0096 & 0.0144 & 0.0126 & 0.0180 & 0.0120 & 0.0119 & 0.0186 & 0.0095 & 0.0121 \\ 
6 & -0.23 & 0.0130 & 0.0209 & 0.0176 & 0.0248 & 0.0189 & 0.0181 & 0.0297 & 0.0160 & 0.0180 \\ 
7 & -0.09 & 0.0194 & 0.0279 & 0.0255 & 0.0343 & 0.0299 & 0.0278 & 0.0381 & 0.0226 & 0.0290 \\ 
8 & 0.05 & 0.0278 & 0.0365 & 0.0368 & 0.0423 & 0.0396 & 0.0388 & 0.0503 & 0.0335 & 0.0390 \\ 
9 & 0.19 & 0.0371 & 0.0423 & 0.0461 & 0.0506 & 0.0495 & 0.0486 & 0.0603 & 0.0395 & 0.0525 \\ 
10 & 0.33 & 0.0447 & 0.0506 & 0.0559 & 0.0571 & 0.0593 & 0.0596 & 0.0705 & 0.0484 & 0.0631 \\ 
11 & 0.47 & 0.0508 & 0.0557 & 0.0630 & 0.0692 & 0.0675 & 0.0680 & 0.0790 & 0.0592 & 0.0755 \\ 
12 & 0.61 & 0.0608 & 0.0630 & 0.0701 & 0.0753 & 0.0784 & 0.0758 & 0.0876 & 0.0641 & 0.0863 \\ 
13 & 0.75 & 0.0658 & 0.0669 & 0.0750 & 0.0816 & 0.0866 & 0.0832 & 0.0940 & 0.0746 & 0.0874 \\ 
14 & 0.89 & 0.0737 & 0.0746 & 0.0855 & 0.0876 & 0.0937 & 0.0907 & 0.0990 & 0.0772 & 0.1025 \\ 
15 & 1.03 & 0.0760 & 0.0769 & 0.0910 & 0.0940 & 0.1011 & 0.0949 & 0.1056 & 0.0838 & 0.1107 \\ 
16 & 1.17 & 0.0858 & 0.0826 & 0.0939 & 0.0950 & 0.1035 & 0.1035 & 0.1113 & 0.0899 & 0.1204 \\ 
17 & 1.31 & 0.0872 & 0.0831 & 0.1026 & 0.1014 & 0.1079 & 0.1067 & 0.1131 & 0.0913 & 0.1252 \\ 
18 & 1.45 & 0.0949 & 0.0898 & 0.1099 & 0.1055 & 0.1184 & 0.1151 & 0.1206 & 0.1012 & 0.1361 \\ 
19 & 1.59 & 0.0964 & 0.0940 & 0.1145 & 0.1108 & 0.1191 & 0.1212 & 0.1286 & 0.1048 & 0.1389 \\ 
20 & 1.73 & 0.1024 & 0.0973 & 0.1192 & 0.1134 & 0.1264 & 0.1249 & 0.1302 & 0.1105 & 0.1470 \\ 
21 & 1.87 & 0.1000 & 0.1004 & 0.1207 & 0.1174 & 0.1288 & 0.1254 & 0.1336 & 0.1111 & 0.1525 \\ 
22 & 2.01 & 0.1029 & 0.0965 & 0.1182 & 0.1124 & 0.1253 & 0.1218 & 0.1331 & 0.1085 & 0.1500 \\ 
23 & 2.15 & 0.0989 & 0.0928 & 0.1122 & 0.1072 & 0.1148 & 0.1146 & 0.1160 & 0.1029 & 0.1458 \\ 
24 & 2.29 & 0.0853 & 0.0788 & 0.0979 & 0.0890 & 0.0998 & 0.0914 & 0.0906 & 0.0888 & 0.1295 \\ 
25 & 2.43 & 0.0618 & 0.0539 & 0.0638 & 0.0538 & 0.0596 & 0.0560 & 0.0548 & 0.0578 & 0.0891 \\ 

\hline
\end{tabular}
\caption{\textbf{Detection efficiencies for the analysed fields.} There are 25 bins equally spaced in $\log\tE$ between $-1.0$ and $2.5$.}
\label{tab:eff}
\end{table}


\section*{References.}

\begin{thebibliography}{1}

\biba{rasio}{Rasio, F.~A. \& Ford, E.~B.}{Dynamical instabilities and the formation of extrasolar planetary systems}{Science}{274}{954-956}{1996}
\biba{marzari}{{Weidenschilling}, S.~J. \& {Marzari}, F.}{Gravitational scattering as a possible origin for giant planets at small stellar distances}{Nature}{384}{619-621}{1996}
\biba{veras}{{Veras}, D. \& {Raymond}, S.~N.}{Planet-planet scattering alone cannot explain the free-floating planet population}{\mnras}{421}{L117-L121}{2012}
\biba{kroupa2003}{Luhman, K.~L.}{The formation and early evolution of low-mass stars and brown dwarfs}{Annu. Rev. Astron. Astrophys.}{50}{65-106}{2012}
\biba{zapatero}{{Zapatero Osorio}, M.~R. \etal}{Discovery of Young, Isolated Planetary Mass Objects in the {$\sigma$} Orionis Star Cluster}{Science}{290}{103-107}{2000}
\biba{liu}{Liu, M.~C. \etal}{The Extremely Red, Young L Dwarf PSO J318.5338-22.8603: A Free-floating Planetary-mass Analog to Directly Imaged Young Gas-giant Planets}{\apjl}{777}{L20}{2013}
\biba{dupuy}{Dupuy, T.~J. \& Kraus, A.~L.}{Distances, Luminosities, and Temperatures of the Coldest Known Substellar Objects}{Science}{341}{1492-1495}{2013}
\biba{scholz}{{Scholz}, A. \etal}{Substellar Objects in Nearby Young Clusters (SONYC). VI. The Planetary-mass Domain of NGC 1333}{\apj}{756}{24}{2012}
\biba{pena}{{Pe{\~n}a Ram{\'{\i}}rez}, K., {B{\'e}jar}, V.~J.~S., {Zapatero Osorio}, M.~R., {Petr-Gotzens}, M.~G. \& {Mart{\'{\i}}n}, E.~L.}{New Isolated Planetary-mass Objects and the Stellar and Substellar Mass Function of the {$\sigma$} Orionis Cluster}{\apj}{754}{30}{2012}
\biba{muzic}{{Mu{\v z}i{\'c}}, K., {Scholz}, A., {Geers}, V.~C. \& 
	{Jayawardhana}, R.}{Substellar Objects in Nearby Young Clusters (SONYC) IX: The Planetary-Mass Domain of Chamaeleon-I and Updated Mass Function in Lupus-3}{\apj}{810}{159}{2015}

\biba{sumi}{Sumi, T. \etal}{Unbound or distant planetary mass population detected by gravitational microlensing}{Nature}{473}{349-352}{2011}
\biba{mao}{{Ma}, S., {Mao}, S., {Ida}, S., {Zhu}, W. \& {Lin}, D.~N.~C.}{Free-floating planets from core accretion theory: microlensing predictions}{\mnras}{461}{L107-L111}{2016}
\biba{udalski}{{Udalski}, A., {Szyma{\'n}ski}, M.~K. \& {Szyma{\'n}ski}, G.}{OGLE-IV: Fourth Phase of the Optical Gravitational Lensing Experiment}{\actaa}{65}{1-38}{2015}
\biba{wozniak}{{Wo{\'z}niak}, P. \& {Paczy{\'n}ski}, B.}{Microlensing of Blended Stellar Images}{\apj}{487}{55-60}{1997}
\biba{bennet}{Bennett, D.~P., Sumi, T., Bond, I.~A., \etal}{Planetary and Other Short Binary Microlensing Events from the MOA Short-event Analysis}{\apj}{757}{119}{2012}
\biba{calchi}{{Calchi Novati}, S., {de Luca}, F., {Jetzer}, P., {Mancini}, L. \& {Scarpetta}, G.}{Microlensing constraints on the Galactic bulge initial mass function}{\aap}{480}{723-733}{2008}
\biba{han1995}{{Han}, C. \& {Gould}, A.}{The Mass Spectrum of MACHOs from Parallax Measurements}{\apj}{447}{53}{1995}
\biba{han2003}{{Han}, C. \& {Gould}, A.}{Stellar Contribution to the Galactic Bulge Microlensing Optical Depth}{\apj}{592}{172-175}{2003}
\biba{lafreniere2007}{{Lafreni{\`e}re}, D. \etal}{The Gemini Deep Planet Survey}{\apj}{670}{1367-1390}{2007}
\biba{bowler2015}{{Bowler}, B.~P., {Liu}, M.~C., {Shkolnik}, E.~L. \& {Tamura}, M.}{Planets around Low-mass Stars (PALMS). IV. The Outer Architecture of M Dwarf Planetary Systems}{\apjs}{216}{7}{2015}
\biba{clanton}{{Clanton}, C. \& {Gaudi}, B.~S.}{Constraining the Frequency of Free-floating Planets from a Synthesis of Microlensing, Radial Velocity, and Direct Imaging Survey Results}{\apj}{834}{46}{2017}
\biba{ida}{Ida, S., Lin., D.~N.~C., \& Nagasawa, M.}{Toward a deterministic model of planetary formation. VII. Eccentricity distribution of gas giants}{\apj}{775}{42}{2013}
\biba{pfyffer}{{Pfyffer}, S., {Alibert}, Y., {Benz}, W. \& {Swoboda}, D.}{Theoretical models of planetary system formation. II. Post-formation evolution}{\aap}{579}{A37}{2015}
\biba{barclay}{Barclay, T., Quintana, E.~V., Raymond, S.~N. \& Penny, M.~T.}{The demographics of rocky free-floating planets and their detectability by WFIRST}{\apj}{841}{86}{2017}
\bibitem{spergel}Spergel, D. \etal Wide-Field InfrarRed Survey Telescope-Astrophysics Focused Telescope Assets WFIRST-AFTA 2015 Report, arXiv:1503.03757 (2015).
\biba{penny}{Penny, M.~T. \etal}{ExELS: an exoplanet legacy science proposal for the ESA Euclid mission - I. Cold exoplanets}{\mnras}{434}{2-22}{2013}
\biba{mao1996}{{Mao}, S. \& {Paczynski}, B.}{Mass Determination with Gravitational Microlensing}{\apj}{473}{57}{1996}
\newcounter{enumTemp}
\setcounter{enumTemp}{\theenumiv}
\end{thebibliography}

\begin{thebibliography}{1}
\setcounter{enumiv}{\value{enumTemp}}
\biba{alard}{{Alard}, C. \& {Lupton}, R.~H.}{A Method for Optimal Image Subtraction}{\apj}{503}{325-331}{1998}
\biba{wozniakdia}{{Wo{\'z}niak}, P.~R.}{Difference Image Analysis of the OGLE-II Bulge Data. I. The Method}{\actaa}{50}{421-450}{2000}
\biba{udalski2003}{{Udalski}, A.}{The Optical Gravitational Lensing Experiment. Real Time Data Analysis Systems in the OGLE-III Survey}{\actaa}{53}{291-305}{2003}
\biba{skowron2016}{{Skowron}, J. \etal}{Analysis of Photometric Uncertainties in the OGLE-IV Galactic Bulge Microlensing Survey Data}{\actaa}{66}{1-14}{2016}
\biba{wyrzykowski2015}{{Wyrzykowski}, {\L}. \etal}{OGLE-III Microlensing Events and the Structure of the Galactic Bulge}{\apjs}{216}{12}{2015}
\biba{eyer}{{Wray}, J.~J., {Eyer}, L. \& {Paczy{\'n}ski}, B.}{OGLE small-amplitude variables in the Galactic bar}{\mnras}{349}{1059-1068}{2004}
\biba{park}{Park, B.-G. \etal}{MOA-2003-BLG-37: A bulge jerk-parallax microlens degeneracy}{\apj}{609}{166-172}{2004}
\biba{jiang}{Jiang, G. \etal}{OGLE-2003-BLG-238: Microlensing mass estimate for an isolated star}{\apj}{617}{1307-1315}{2004}
\biba{smith}{Smith, M.~C., Wo{\'z}niak, P., Mao, S. \& Sumi, T.}{Blending in gravitational microlensing experiments: source confusion and related systematics}{\mnras}{380}{805-818}{2007}
\biba{hawley}{Hawley, S.~L. \etal}{Kepler Flares. I. Active and Inactive M Dwarfs}{\apj}{797}{121}{2014}
\biba{holtz1998}{{Holtzman}, J.~A. \etal}{The Luminosity Function and Initial Mass Function in the Galactic Bulge}{\aj}{115}{1946-1957}{1998}
\biba{gould1992}{Gould, A.}{Extending the MACHO search to about $10^6$ solar masses}{\apj}{392}{442}{1992}
\biba{bennett}{Bennett, D.~P. \& Rhie, S.~H.}{Detecting Earth-mass planets with gravitational microlensing}{\apj}{472}{660-664}{1996}
\biba{kiraga1994}{{Kiraga}, M. \& {Paczynski}, B.}{Gravitational microlensing of the Galactic bulge stars}{\apjl}{430}{L101-L104}{1994}
\biba{han1996}{{Han}, C. \& {Gould}, A.}{Statistical Determination of the MACHO Mass Spectrum}{\apj}{467}{540}{1996}
\biba{bissantz2004}{{Bissantz}, N., {Debattista}, V.~P. \& {Gerhard}, O.}{Large-Scale Model of the Milky Way: Stellar Kinematics and the Microlensing Event Timescale Distribution in the Galactic Bulge}{\apjl}{601}{L155-L158}{2004}
\biba{mao2005}{{Wood}, A. \& {Mao}, S.}{Optical depths and time-scale distributions in Galactic microlensing}{\mnras}{362}{945-951}{2005}
\biba{dwek1995}{{Dwek}, E. \etal}{Morphology, near-infrared luminosity, and mass of the Galactic bulge from COBE DIRBE observations}{\apj}{443}{716-730}{1995}
\biba{zheng2001}{{Zheng}, Z., {Flynn}, C., {Gould}, A., {Bahcall}, J.~N. \& {Salim}, S.}{M Dwarfs from Hubble Space Telescope Star Counts}{\apj}{555}{393-404}{2001}
\biba{gould2000}{{Gould}, A.}{Measuring the Remnant Mass Function of the Galactic Bulge}{\apj}{535}{928-931}{2000}
\biba{williams2009}{{Williams}, K.~A., {Bolte}, M. \& {Koester}, D.}{Probing the Lower Mass Limit for Supernova Progenitors and the High-Mass End of the Initial-Final Mass Relation from White Dwarfs in the Open Cluster M35 (NGC 2168)}{\apj}{693}{355-369}{2009}
\biba{kiziltan}{{Kiziltan}, B., {Kottas}, A., {De Yoreo}, M. \& {Thorsett}, S.~E.}{The Neutron Star Mass Distribution}{\apj}{778}{66}{2013}
\biba{ozel2010}{{{\"O}zel}, F., {Psaltis}, D., {Narayan}, R. \& {McClintock}, J.~E.}{The Black Hole Mass Distribution in the Galaxy}{\apj}{725}{1918-1927}{2010}
\biba{zoccali2000}{{Zoccali}, M. \etal}{The Initial Mass Function of the Galactic Bulge down to \~{}0.15 M$_{solar}$}{\apj}{530}{418-428}{2000}
\biba{belczynski2008}{{Belczynski}, K. \etal}{Compact Object Modeling with the StarTrack Population Synthesis Code}{\apjs}{174}{223-260}{2008}
\biba{kroupa2001}{{Kroupa}, P.}{On the variation of the initial mass function}{\mnras}{322}{231-246}{2001}
\biba{wegg2016}{{Wegg}, C., {Gerhard}, O. \& {Portail}, M.}{MOA-II Galactic microlensing constraints: the inner Milky Way has a low dark matter fraction and a near maximal disc}{\mnras}{463}{557-570}{2016}
\biba{oliveira}{{Alves de Oliveira}, C.}{The low mass end of the IMF}{\memsai}{84}{905}{2013}
\biba{allen2005}{{Allen}, P.~R., {Koerner}, D.~W., {Reid}, I.~N. \& {Trilling}, D.~E.}{The Substellar Mass Function: A Bayesian Approach}{\apj}{625}{385-397}{2005}
\biba{quanz2012}{{Quanz}, S.~P., {Lafreni{\`e}re}, D., {Meyer}, M.~R., 
	{Reggiani}, M.~M. \& {Buenzli}, E.}{Direct imaging constraints on planet populations detected by microlensing}{\aap}{541}{A133}{2012}
\end{thebibliography}
\end{document}